\documentclass[]{spie}  

\usepackage{amsmath,amsfonts,amssymb}
\usepackage{graphicx}
\usepackage[colorlinks=true, allcolors=blue]{hyperref}
\usepackage{caption} 
\title{Enhancing Organ at Risk Segmentation with Improved Deep Neural Networks}

\author[a]{Ilkin Isler}
\author[a,b]{Curtis Lisle}
\author[c]{Justin Rineer}
\author[c]{Patrick Kelly}
\author[a]{Damla Turgut}
\author[c]{Jacob Ricci}
\author[a,d]{Ulas Bagci}
\affil[a]{Dept. of Computer Science, University of Central Florida, Orlando, FL, USA}
\affil[b]{KnowledgeVis LLC, Orlando, FL, USA}
\affil[c]{Dept. of Radiation Oncology, Orlando Health, Orlando, FL, USA.}
\affil[d]{Dept. of Radiology and BME, Northwestern University, Chicago, IL, USA.}

\authorinfo{Further author information: (Send correspondence to U.B.)\\U.B: E-mail: ulas.bagci@northwestern.edu}

\begin{document} 
\maketitle

\begin{abstract}
Organ at risk (OAR) segmentation is a crucial step for treatment planning and outcome determination in radiotherapy treatments of cancer patients. Several deep learning based segmentation algorithms have been developed in recent years, however, U-Net remains the de facto algorithm designed specifically for biomedical image segmentation and has spawned many variants with known weaknesses. In this study, our goal is to present simple architectural changes in U-Net to improve its accuracy and generalization properties. Unlike many other available studies evaluating their algorithms on single center data, we thoroughly evaluate several variations of U-Net as well as our proposed enhanced architecture on multiple data sets for an extensive and reliable study of the OAR segmentation problem. Our enhanced segmentation model includes (a)architectural changes in the loss function, (b)optimization framework, and (c)convolution type.  Testing on three publicly available multi-object segmentation data sets, we achieved an average of 80\% dice score compared to the baseline U-Net performance of 63\%. 

\end{abstract}


\section{INTRODUCTION}
\label{sec:intro}  
Cancer is one of the leading causes of death. In 2021, there will be an estimated 1.9 million new total cancer cases diagnosed and 608,570 total cancer deaths in the United States~\cite{gansler2010sixty}. 
Radiation therapy is the treatment of cancer using targeted beams of radiation applied to cancerous tissues. During this process, non-cancerous tissue is inevitably exposed to damaging radiation. 
An accurate medical image segmentation of the tumor(s) and the surrounding organs/tissues, called organ at risk (OAR) segmentation, is crucial for radiation therapy planning. Although medical image segmentation has long been studied, the current clinical routine still includes tedious and highly variable manual image segmentation for treatment planning. Introduced by Ronneberger et al.~\cite{ronneberger2015u} and based on fully convolutional networks~\cite{long2015fully} (FCNs), U-Net is the most used deep learning based segmentation algorithm in medical imaging in general, and OAR in particular. 
Due to its flexibility to be adapted to a variety of such tasks, U-Net has become one of the de facto segmentation models in the current literature, with many variants developed to improve upon it that have achieved excellent results~\cite{cciccek20163d,kerfoot2018left,zhou2018unet++}. However, the field is still open to improvement due to the suboptimality and weaknesses of existing U-Net solutions. \textbf{Our goals} in this paper are to evaluate the generalizability of multi-class segmentation methods across multiple data sets (lung and head/neck cancers) and propose simple changes to the baseline deep segmentation architectures that enhance segmentation results in both accuracy and generalizability. 

\section{METHODS}
\subsection{Segmentation Models}

We evaluated the performance of several U-Net variants. The models we used are U-Net with 2D images (U-Net), U-Net with residual units with 2D and 3D images (ResU-Net and 3D ResU-Net), ResU-Net with dilated convolutions with 2D images, and U-Net++ with 2D images (U-Net++).  \textbf{U-Net} is a U-shaped architecture with an encoder used to capture image context and a decoder enabling precise local information with skip connections from each down layer to each corresponding up layer. \textbf{ResU-Net} is a U-Net convolutional neural network architecture that utilizes skip connections and also residual units. The residual units feed forward the feature map from one layer to a deeper layer in the network. This helps reduce the vanishing gradient problem and generally provides improved stability and performance for deeper networks. \textbf{ResU-Net} with residual units and dilated convolutions expands the receptive field of the neural network, better preserving spatial information. \textbf{U-Net++} is a deeply-supervised encoder-decoder network where the encoder and decoder sub-networks are connected through a series of nested, dense skip pathways.



\subsection{Architectural Optimizations and Evaluation Metrics}
We utilized two different conventional metrics for evaluation: DICE and Hausdorff Distance (HD). The DICE coefficient is broadly used in medical image segmentation, indicating the region-wise similarities (overlaps) between two segmented objects. Hausdorff distance is a metric for calculating the dissimilarity between the boundaries of two segmented objects. For architectural optimization, we made simple yet effective changes that include differences in loss functions, learning rate schedulers, encoder type, convolution type, and normalization to achieve better segmentation results. 

\textbf{Loss Function:} We used two loss functions for our experiments; namely DICE loss and a weighted combination of DICE and Cross Entropy (CE) loss. When it comes to small objects, using DICE loss alone results in lower accuracy. In cases where the predicted area does not overlap with the label region, we still want the prediction to be as close as possible to the label, which is where CE loss helps.

\textbf{Optimizer and Scheduler:} We investigated the effect of using the Cyclic Learning Rate (CyclicLR) scheduler~\cite{smith2017cyclical} with the widely used ADAM optimizer. With CyclicLR, the learning rate varies cyclically between the base learning rate and a given maximum value instead of monotonically decreasing. This leads to improved results with fewer tuning requirements and often results in convergence in fewer iterations. We chose 0.001 and 0.006 as the base and maximum learning rates, respectively. 

\textbf{Encoder Type:} We experimented on the effect of two different encoders: ResNet34 and EfficientNet-b4. The EfficientNet, as proposed in Ref.~\citenum{tan2019efficientnet}, consists of compound coefficients that define model scaling and adjust the depth, width, and resolution of the network for better performance. 

\textbf{Convolution Type:} We examined the effect of dilated convolutions. Dilated convolutions enlarge the receptive field without losing resolution or exponentially increasing the number of parameters. In our experiments, we used a dilation factor of 3. 

\textbf{Normalization:} Instance normalization, is applied after the convolutions in our ResU-Nets (ResU-Net, 3D ResU-Net, and dilated ResU-Net).

\section{EXPERIMENTS and RESULTS}
\subsection{Datasets and Preprocessing}
We evaluated the models on two different head and neck cancer data sets and one lung cancer data set.  These data sets include both small, medium, and large objects to test the generalizability of our models.

\textbf{OpenKBP}~\cite{babier2020openkbp}: The Open-Access Knowledge-Based Planning Grand Challenge's data set contains 128x128x128 3D images from 340 patients who were treated for head-and-neck cancer with intensity modulated radiation therapy that was contoured by clinicians at twelve institutions with different planning protocols. Of the 340 patients, we selected the greatest number of common OARs, resulting in 188 patients who had available contours for five OARs: brainstem, spinal cord, right parotid, left parotid, and mandible. We merged the individual masks together for our multi-class segmentation tasks.
 
\textbf{PDDCA (A Public Domain Database for Computational Anatomy)}~\cite{raudaschl2017evaluation}:  The Public MICCAI 2015 dataset, used for a segmentation challenge contains 48 patients images with 9 different OAR annotations of size 512x512xZ, where z is the depth of the patient’s scan and varies. We selected patients who had all of the six OARs we chose to be similar to the OpenKBP data: brainstem, chiasm, optic nerve left, optic nerve right, parotid left, parotid right.

\textbf{NSCLC-Radiomics}~\cite{aerts2014decoding}: This collection contains images from 422 non-small cell lung cancer (NSCLC) patients and their manual delineation (primary gross tumor volume ("GTV-1") and selected anatomical structures (i.e., lung, heart and esophagus)) by a radiation oncologist. After extracting DICOM format imagery from a publicly available dataset from The Cancer Imaging Archive (TCIA)~\cite{clark2013cancer} and downloading the data, we chose 306 patients who have all three of the spinal cord, left lung, and right lung masks.

All data sets were reformatted into NumPy tensors for deep learning at native resolution without resampling. For data augmentation, we applied well-known random transforms, using the emerging MONAI medical imaging framework~\cite{MONAI_Consortium_MONAI_Medical_Open_2020}: boundary cropping, intensity normalization, contrast adjustment, affine transformation, 3D elastic transformation, and random Gaussian noise.  We partitioned the data sets into train, validation, and test with the ratios of 0.7, 0.15, 0.15, respectively. The results presented in this paper are test results.

\subsection{Enhancing Segmentation Efficacy}
As mentioned in the nnU-Net paper~\cite{isensee2018nnu}, many of the previous works seem to achieve better performance with architectural tweaks although the benchmark (U-Net) they’re using is not even a fully optimized network. The study claims that to see the real performance of a network and in this case the benchmark, first, the network should be fully adapted to that specific task. According to the preliminary experiments, when some non-architectural changes are applied to optimize the network, the architectural tweaks being made on the fully optimized network are unable to improve segmentation results and thus most likely unable to advance the state-of-the-art. In the light of these findings, in this paper, we’re showing how the little but efficient tweaks can increase the performance of our network. 

\begin{table}[hbt!]
\centering
\caption{Evaluating the performance of several U-Net variants (U-Net with 2D images (U-Net), U-Net++ with 2D images (U-Net++), U-Net with residual units with 2D and 3D images (ResU-Net and 3D ResU-Net) and ResU-Net with dilated convolutions with 2D images) on NSCLC (contains left and right lung and spinal cord) dataset for 150 epochs. DICE scores are being showed.}
\label{fig:variants}
\begin{tabular}{|l|l|l|l|l|l|}
\hline
NSCLC(150)  & U-Net & U-Net++ & ResU-Net  & Dilated ResU-Net  & 3D ResU-Net  \\ 
\cline{1-6}
Lung R          & 0.73   & 0.74  & 0.72  & 0.72 & 0.94  \\ 
\hline
Lung L          & 0.72   & 0.73  & 0.71  & 0.71  & 0.93  \\ 
\hline
Spinal Cord     & 0.83  & 0.84  & 0.52  & 0.52 & 0.77  \\ 
\hline
\textbf{Overall}    &\textbf{0.76} &\textbf{0.77} &\textbf{0.65}&\textbf{0.65} &\textbf{0.88}\\
\hline
\end{tabular}
\end{table}

\begin{figure}[hbt!]
    \centering
    \includegraphics[width=10cm]{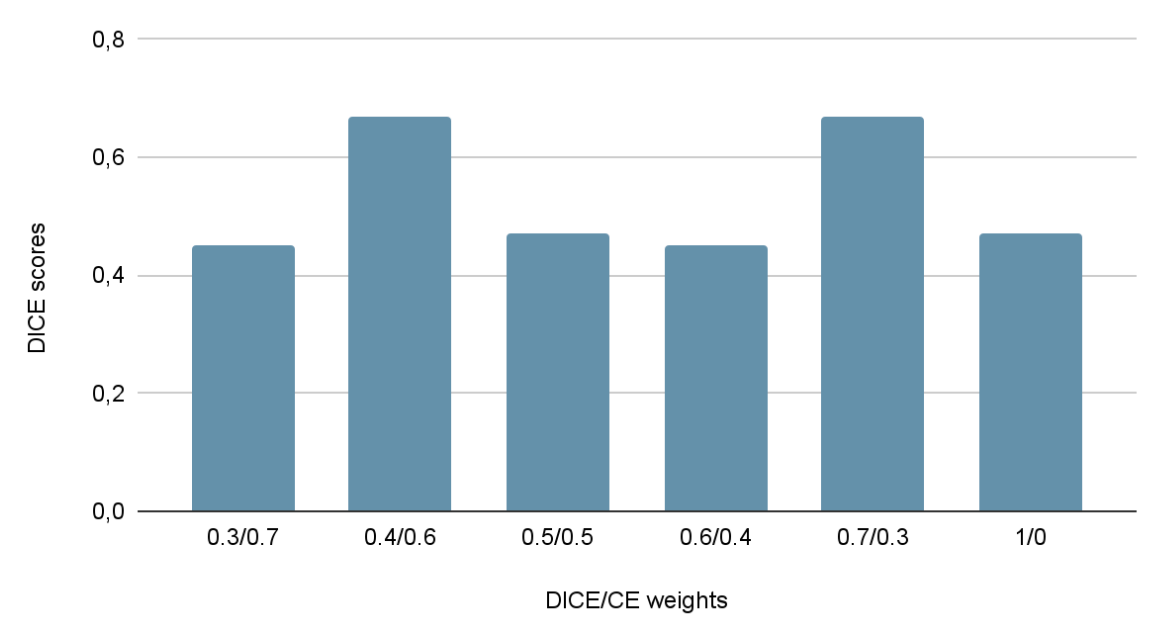}
    \caption{Comparing different DICE and CE weights for the loss function with the enhanced model on H\&N dataset: PDDCA.}
  \label{fig:diceceloss}
\end{figure}

\begin{figure}[hbt!]
    \centering
    \includegraphics[width=17cm]{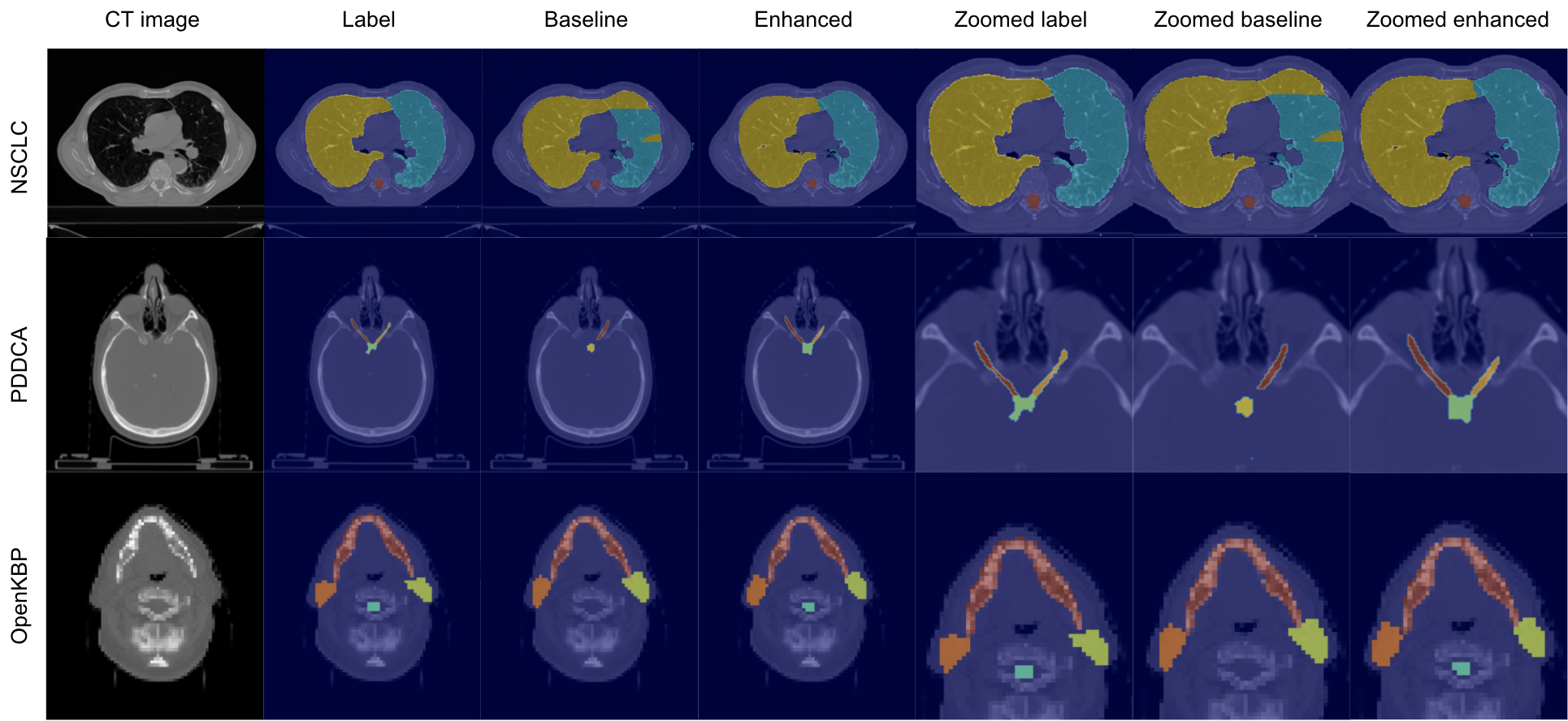}
    \caption{Comparing the baseline model with the enhanced model on lung and H\&N datasets: NSCLC, OpenKBP and PDDCA.}
    \label{fig:results}
\end{figure}

We started with comparing several U-Net variants as shown in Table \ref{fig:variants}. Between the variants, 3D ResU-Net was the one that performed the best on the NSCLC dataset. The improvement between U-Net and U-Net++ was too small. Also, changing the convolutional dilation parameter to 3 for the dilated ResU-Net didn't seem to improve the ResU-Net model. Even though we trained all the 2D models with all the slices, the 3D model (3D ResU-Net) achieved greater performance since it was able to learn the relations between the slices better. Note that the 3D Res-UNet used in this table is the same as the baseline in \ref{fig:nsclc}. 
We experimented using different weight combinations for the terms in the combined DICE-CE loss. Optimal accuracy was achieved with the weighting of 0.4 and 0.6, respectively, for the DICE and CE losses as shown in Figure \ref{fig:diceceloss}.  Once we found the combination of weights that yielded the highest DICE testing score, we incorporated this loss function during training. 
We trained the 3D ResU-Net on the PDDCA dataset with three built-in cyclic learning rate policies: '\textbf{triangular}', '\textbf{triangular2}', and '\textbf{exp\_range}', as detailed in the original paper. In CyclicLR, we change the learning rate (lr) between a lower and higher threshold. The default one, triangular, linearly increases and decreases the lr between the maximum lr and the base lr at each cycle. For triangular2, the maximum lr is halved after every cycle. For exp\_range,  maximum lr is reduced exponentially with each iteration.
 The best DICE scores were received while using 'exp\_range' as shown in Table \ref{fig:cyclic}. Using 'exp\_range' yielded to 59\% improvement on the DICE score.
Comparing the baseline results with the enhanced version where we replaced the DICE Loss with DICE-CE Loss and used CyclicLR in addition, we achieved considerable improvement especially on small objects which are more difficult to segment as shown in Table \ref{fig:openkbp}, \ref{fig:pddca} and \ref{fig:nsclc} and Figure \ref{fig:results}. Enhanced models improved DICE and HD95 scores and converged faster on all datasets.

\begin{table}[hbt!]
\centering
\caption{Comparing the effect of different Cyclic Learning Rate schedulers (triangular, triangular2 and exp\_range) on DICE scores on PDDCA dataset with 3D ResU-Net for 300 epochs.}
\begin{tabular}{|l|l|l|l|l|} 
\hline
PDDCA(300)    & without cyclicLR    & triangular  & triangular2   & exp\_range \\ 
\hline
Brainstem     & 0.59    & 0.83    & 0.59      & 0.83 \\ 
\hline
Chiasm       & 0.34 &  0.52     & 0.31      & 0.52 \\ 
\hline
Optic Nerve L & 0.31    &  0.58    & 0.37      & 0.58 \\ 
\hline
Optic Nerve R & 0.22    & 0.54    & 0.33      & 0.54 \\ 
\hline
Parotid L     & 0.55    & 0.79    & 0.62      & 0.79  \\  
\hline
Parotid R     & 0.53    & 0.76    & 0.59      & 0.76  \\ 
\hline
\textbf{Overall}    &\textbf{0.42}  &\textbf{0.67}    &\textbf{0.47}      &\textbf{0.67}\\
\hline
\end{tabular}
\label{fig:cyclic}
\end{table}

\begin{table}[hbt!]
\parbox{.45\linewidth}{
\centering
\caption{Comparing the baseline model with the enhanced model on OpenKBP dataset with 3D ResU-Net for 300 epochs. The HD95 scores are in millimeters.}
\begin{tabular}{|l|l|l|l|l|} 
\hline
3D ResU-Net   & \multicolumn{2}{l|}{Baseline} & \multicolumn{2}{l|}{Enhanced}  \\ 
\cline{2-5}
OpenKBP(300)    & DICE  & HD95        & DICE  & HD95     \\ 
\hline
Brainstem       &0.52     &2.19       &0.80   &3.94\\ 
\hline
Spinal Cord     &0.50      &5.77       &0.75    &5.97\\ 
\hline
Parotid R       &0.73       &2.62       &0.76  &2.31\\ 
\hline
Parotid L       &0.75       &2.51       &0.75  &2.28\\ 
\hline
Mandible        &0.84       &2.12       &0.86   &1.78\\ 
\hline
\textbf{Overall} &\textbf{0.67}  &\textbf{3.04} &\textbf{0.78}  &\textbf{3.26}\\
\hline
\end{tabular}
\label{fig:openkbp}
}
\parbox{.60\linewidth}{
\centering
\parbox{.80\linewidth}{
\centering
\caption{Comparing the baseline model with the enhanced model on PDDCA dataset with 3D ResU-Net for 300 epochs. The HD95 scores are in millimeters.}
\label{fig:pddca}
}
\begin{tabular}{|l|l|l|l|l|} 
\hline
3D ResU-Net   & \multicolumn{2}{l|}{Baseline} & \multicolumn{2}{l|}{Enhanced}  \\ 
\cline{2-5}
PDDCA(300)    & DICE  & HD95   & DICE & HD95      \\ 
\hline
Brainstem       &0.57     &45.96      &0.81   &33.00\\ 
\hline
Chiasm          &0.30    &119.55     &0.53  &24.48\\ 
\hline
Optic Nerve L   &0.15    &172.69     &0.63  &16.83\\ 
\hline
Optic Nerve R   &0.04    &185.80        &0.57      &16.56\\ 
\hline
Parotid R       &0.53   &116.20     &0.81  &26.50\\ 
\hline
Parotid L        &0.51  &59.83      &0.79  &7.36\\ 
\hline
\textbf{Overall} &\textbf{0.35}  &\textbf{116.67}& \textbf{0.69} &\textbf{20.79}\\
\hline
\end{tabular}
}
\end{table}

\begin{table}[hbt!]
\centering
\caption{Comparing the baseline model with the enhanced model on NSCLC dataset with 3D ResU-Net for 150 epochs. The HD95 scores are in millimeters.}
\label{fig:nsclc}
\begin{tabular}{|l|l|l|l|l|}
\hline
3D ResU-Net   & \multicolumn{2}{l|}{Baseline} & \multicolumn{2}{l|}{Enhanced}  \\ 
\cline{2-5}
NSCLC(150)    & DICE & HD95(mm)   & DICE & HD95 (mm)      \\ 
\hline
Lung R          & 0.94   & 17.93  & 0.97  & 4.16  \\ 
\hline
Lung L          & 0.93   & 36.52  & 0.97  & 4.47  \\ 
\hline
Spinal Cord     & 0.77  & 89.37  & 0.84  & 11.82  \\ 
\hline
\textbf{Overall}    &\textbf{0.88} &\textbf{47.94} &\textbf{0.93} &\textbf{6.82}\\
\hline
\end{tabular}
\end{table}

For the NSCLC-Radiomics (lungs and spinal cord) dataset, training both models (baseline and enhanced) for 150 epochs was enough for the models to converge. Even though the model easily learned how to segment the lungs since they are relatively large, it was harder for the model to learn how to segment the spinal cord. For the H\&N datasets (OpenKBP and PDDCA), we trained both models for 300 epochs since there are small, hard to learn organs in these datasets such as chiasm and optic nerves.

\begin{table}[hbt!]
\centering
\caption{Comparing the effect of different encoders: efficientnet-b4 and resnet34 Cyclic Learning Rate schedulers (triangular, triangular2 and exp\_range) on DICE scores on PDDCA dataset with 3D ResU-Net for 300 epochs.}
\begin{tabular}{|l|l|l|} 
\hline
PDDCA(300)    & efficientnet    & resnet34 \\ 
\hline
\textbf{Overall}    &\textbf{0.67}  &\textbf{0.45} \\
\hline
\end{tabular}
\label{fig:encoder}
\end{table}

\begin{figure}[hbt!]
    \centering
    \includegraphics[width=17cm]{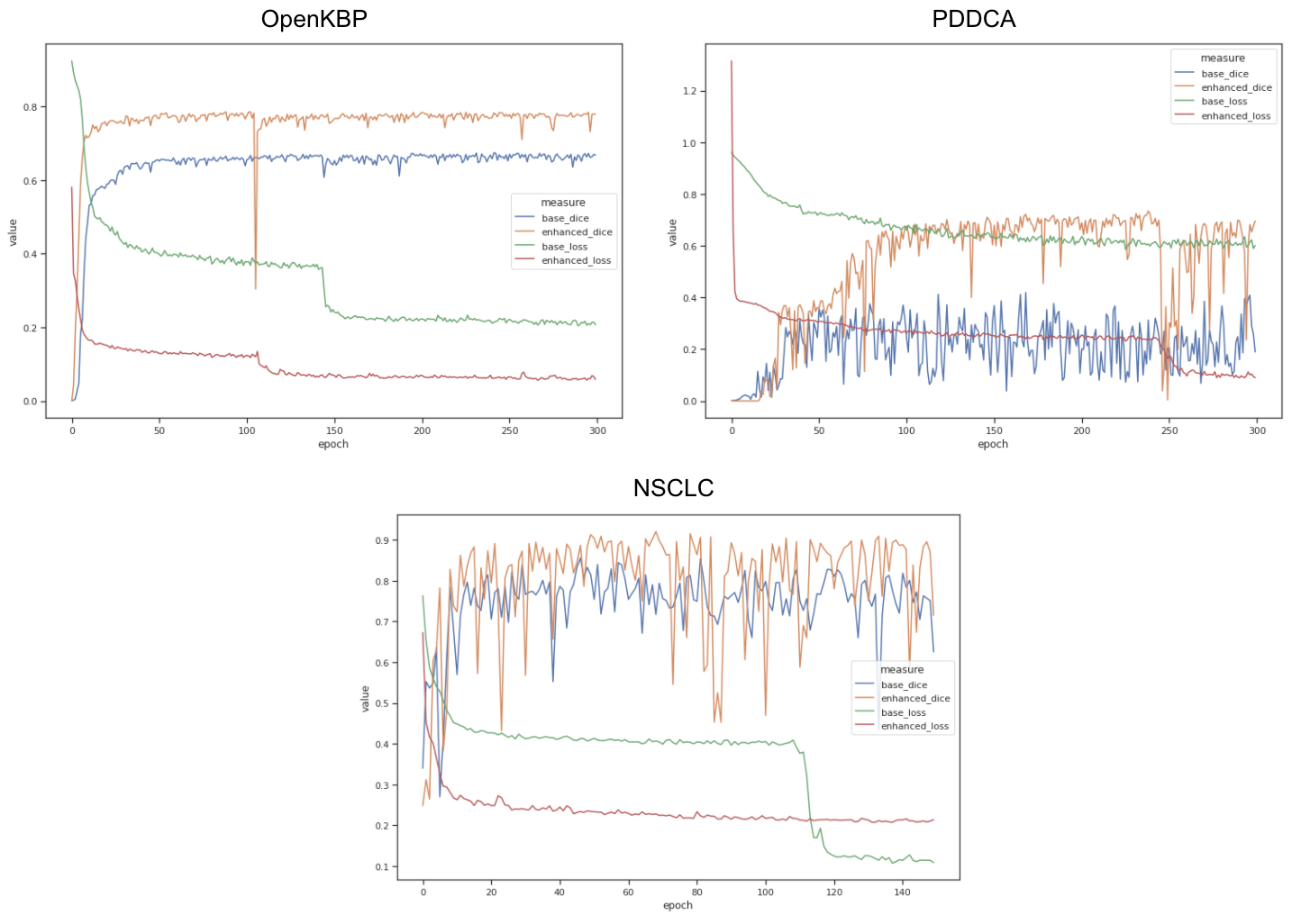}
    \caption{DICE and loss graphs for both enhanced and baseline models for all 3 data sets.}
  \label{fig:graphs}
\end{figure}

Applying different encoders to the enhanced version showed us that using EfficientNet as the encoder gives better accuracy as shown in Table \ref{fig:encoder}. For all the 3 datasets, our enhanced version improved both DICE scores and HD95 scores. The improvements on the DICE scores are 16\%, 97\%, and 5\% for the OpenKBP, PDDCA, and NSCLC datasets, respectively. In Figure \ref{fig:graphs}, we notice the fluctuating of dice score lines which are caused by cyclic learning which actually helps with fast convergence. Depending on the dataset, whether it has hard to learn organs or not, the speed of convergence varies. For all data sets, we can clearly see that the enhanced versions perform better.

\section{CONCLUSIONS AND FUTURE WORK}
In this study, we thoroughly evaluated several variations of U-Net as well as our proposed enhanced architecture on multiple data sets for an extensive evaluation of model performance in OAR segmentation. Our enhanced segmentation model includes architectural changes in the loss function, optimization technique,  and convolution type that substantially improved accuracy while still delivering effective training on data sets containing organs of different sizes. 

As future work, we are currently evaluating our models on an additional radiotherapy treatment planning data set from a local partner hospital as part of a joint project to develop novel therapy tools. This additional data set will allow us to further evaluate the generalizability of our model enhancement approach.

\section*{ACKNOWLEDGMENTS} 
This study greatly acknowledges the funding source: Florida Dept of Health (FDOH)-20K04. Dr. Bagci acknowledges the partial support by the NIH grants R01-CA246704 and R01-CA240639. 

\bibliography{report} 
\bibliographystyle{spiebib} 

\end{document}